\documentclass[runningheads]{llncs}
\usepackage[T1]{fontenc}
\usepackage{graphicx}
\usepackage{booktabs}

\usepackage{array}
\newcolumntype{P}[1]{>{\centering\arraybackslash}p{#1}}

\usepackage{xcolor}
\definecolor{darkgreen}{rgb}{0.01, 0.75, 0.24}

\usepackage{amssymb}
\usepackage{pifont}

\makeatletter
\newcommand{\setlabel}[1]{\edef\@currentlabel{#1}\label}
\makeatother

\newcommand{\X}{\textcolor{darkgreen}{\ding{51}}}
\newcommand{\til}{\textcolor{orange}{\huge\textbf{\raisebox{0.2ex}{\texttildelow}}}}

\newcommand{\low}{\textcolor{darkgreen}{\textbf{1 - Low}}}
\newcommand{\medium}{\textcolor{orange}{\textbf{2 - Medium}}}
\newcommand{\high}{\textcolor{red}{\textbf{3 - High}}}

\newcommand{\zerotwo}{\textcolor{darkgreen}{\textbf{0.2 - Low}}}
\newcommand{\zerofive}{\textcolor{orange}{\textbf{0.5 - Medium}}}
\newcommand{\zeronine}{\textcolor{red}{\textbf{0.9 - High}}}

\newcommand{\overallmedium}{\textcolor{orange}{\textbf{0.8 - Medium}}}
\newcommand{\overallhigh}{\textcolor{red}{\textbf{2.4 - High}}}

\usepackage{hyperref}
\usepackage{cleveref}

\usepackage{amsthm}
\theoremstyle{definition}
\newtheorem{scen}{Use Case}[]

\crefname{scen}{Use Case}{Use Cases}

\newtheorem{att}{Attack}[scen]
\crefname{att}{Attack}{Attacks}

\newtheorem{an}{First Analysis of Attack}[scen]

\newtheorem{prop}{Proposition}
\crefname{prop}{Proposition}{Propositions}

\newtheorem{prope}{Property}
\crefname{prope}{Property}{Properties}

\usepackage{todonotes}

\begin{document}
\title{Cybersecurity and Embodiment Integrity for Modern Robots: A Conceptual Framework}
\titlerunning{Cybersecurity and Embodiment Integrity for Modern Robots}
\author{Alberto Giaretta \and
Amy Loutfi}

\authorrunning{Giaretta and Loutfi}
\institute{Department of Computer Science, Örebro University, Sweden \email{\{alberto.giaretta,amy.loutfi\}@oru.se}}
\maketitle              % typeset the header of the contribution
\begin{abstract}
Thanks to new technologies and communication paradigms, such as the Internet of Things (IoT) and the Robotic Operating System (ROS), modern robots can be built by combining heterogeneous standard devices in a single embodiment. Although this approach brings high degrees of modularity, it also yields uncertainty, with regard to providing cybersecurity assurances and guarantees on the integrity of the embodiment. In this paper, first we illustrate how cyberattacks on different devices can have radically different consequences on the robot's ability to complete its tasks and preserve its embodiment. We also claim that modern robots should have self-awareness for what concerns such aspects, and formulate in two propositions the different characteristics that robots should integrate for doing so. Then, we show how these propositions relate to two established cybersecurity frameworks, the NIST Cybersecurity Framework and the MITRE ATT\&CK, and we argue that achieving these propositions requires that robots possess at least three properties for mapping devices and tasks. Last, we reflect on how these three properties could be achieved in a larger conceptual framework.

\end{abstract}
\section{Introduction}
\label{sec:intro}
Modern robots are more modular than in the past, when robots were monolithic machines, manufactured to perform specific tasks and equipped with sensors and actuators developed ad-hoc. Thanks to new technologies, such as the Internet of Things (IoT) and the Robot Operating System (ROS), engineers can select off-the-shelf IoT devices, combine them in a single embodied physical system, and put them in communication using ROS as a middleware framework. This setup makes it possible to integrate new devices with low time overhead, swap them out, and reconfigure their communication flows. In other words, this allows to create robots that are not necessarily cogent and coherent systems, but systems of interchangeable systems that cooperate to achieve specialised tasks.

Although this ensures high degrees of modularity and interchangeability, it also exacerbates cybersecurity threats. A monolithic robot is expected to have little to no dynamic changes to its devices composition. Sensors and actuators installed on the embodiment might be replaced with exact replicas in case of malfunction, but this does not change the fundamental nature of the robot, nor its functions. This certainty allows the engineer to reason in advance regarding the cybersecurity properties of the devices composing the embodied robot - hereby defined \textit{embodied devices}. In contrast, modular robotic systems could integrate devices from various manufacturers and be reconfigured extensively. This makes it difficult to preemptively reason about their cybersecurity properties.

Besides communicating and cooperating in the virtual space, these devices are physically embodied in a complex system. This entails that a failure in one of the sensors or actuators yields risks for the embodiment as a whole, depending on the embodied devices roles. For example, a cyberattack struck on a camera sensor could lead a robot to drive into a wall, damaging itself and creating hazards for surrounding people. These unprecedented levels of interchangeability, modularity, and embodiment, combined with the requirements for robots to perceive the environment and adapt to changing external conditions, exacerbate the problem of providing cybersecurity assurances. In this paper, we explore the connections between devices, embodiment, and cybersecurity aspects.

%\subsection{Outline}
%This paper is structured as follows. In \Cref{sec:rel_work}, we provide a brief overview of relevant work in the field, concerning cybersecurity in the robotic field, in particular related to ROS, and on robotics embodiment. In \Cref{sec:scenarios}, we present use cases of two robots equipped with some devices and, for each use case, two hypothetical cybersecurity attacks targeted at different devices. In \Cref{sec:propositions}, we analyse in depth the cyberattacks we presented, as well as the consequences of threats on robotic tasks and embodiment. Then, we formalise in two propositions the characteristics that should be integrated in a robotic framework for mitigating the threats. In \Cref{sec:framework}, we describe three base properties that a framework should possess to achieve the two propositions previously mentioned, and we list the main components that make these properties obtainable. Last, in \Cref{sec:conclusion} we tie together our reflections and draw our conclusions, for what it concerns the desirable characteristics and components that robotic frameworks should possess, for achieving a holistic evaluation of cybersecurity properties and embodiment.

\section{Related Work}
\label{sec:rel_work}
In the robotics field, developers have adopted many ad-hoc solutions to develop their robotic frameworks. The high specificity of these solutions makes it hard to use them for wide-range reflections on a unified conceptual framework, which takes into account the correlation between embodiment and cybersecurity. 
Embodied robotics considers the importance of the body in shaping the robot's interactions with its surroundings. Here, the physical embodiment of a robot includes its sensors, actuators, and other components that allow it to sense, move, and manipulate objects in the real world \cite{cangelosi2015embodied}.  Whether the agent is a physical agent, or a simulated one, embodied cognition purports that the embodied instantiation is essential for driving the cognitive processes \cite{chrisley2003embodied}. Pragmatically, embodiment has additional benefits, including the integration of sensorimotor responses, offloading of certain processes to either the body or the environment, and engagement in more natural or intuitive interactions \cite{corti2022robot}. 

The standard tool for implementing modular robotic platforms is the Robot Operating System (ROS)~\cite{quigley2009ros}. ROS is a framework that provides sensors and actuators with a communication layer, which allows for easy development of complex systems using heterogeneous off-the-shelf devices. Despite being the most widely used robotic framework in academia, its lack of cybersecurity guarantees is one of the main reasons why it is not used in commercial and government applications~\cite{mcclean2013preliminary}. Among the vulnerabilities detected over the years in ROS, research teams highlighted plain-text communications, unprotected TCP ports, unencrypted data storage~\cite{mcclean2013preliminary}, as well as data injection and manipulation~\cite{dieber2016application}.

Various strategies for strengthening ROS have been proposed in the literature, and most of these strategies focus on encrypting ROS communication channels. Dieber et al.~\cite{dieber2016application} proposed a security architecture that sits at the application layer on top of ROS, and provides cryptographic methods to ROS, for ensuring data confidentiality and integrity. In 2017, Dieber et al.~\cite{dieber2017security} chose a similar strategy for hardening ROS cybersecurity, proposing an application-based solution which integrates directly into the core of ROS and enables secure communication channels. Also Balsa-Comeròn et al.~\cite{balsa2018cybersecurity} highlighted that, although used by most research institutes, ROS misses some basic security features, such as encrypted communications and rules for ensuring correct behaviour. In their paper, the authors implement ROSRV, a framework that encrypts communications with a standard AES algorithm, and defines semantic rules for ROS messages to enforce expected actions. Their experiments show that the framework successfully prevents plain-text messages from being sent, and detects denial-of-service (DoS) attacks, thanks to its semantic rules. Despite academic proposals and efforts to harden ROS, in 2019 Rivera et al.~\cite{rivera2019rosploit} noted that cybersecurity research on ROS is still insufficient. The authors proposed ROSploit, a penetration testing suite that allows for detecting and exploiting vulnerabilities in ROS systems, hence improving the overall cybersecurity levels of ROS-based robotic implementations. Caiazza et al.~\cite{caiazza2019enhancing} provided a set of measures and best practices that should be followed to design security solutions for ROS-based systems, without negatively affecting their performance. Teixeira et al.~\cite{teixeira2020security} created a tool that shows the capability of intercepting, manipulating, and entirely disrupting regular communications between ROS nodes. 

More recently, Mayoral-Vilches et al.~\cite{mayoral2022sros2} discuss the security properties and shortcomings of ROS 2, a redesign of ROS that builds upon Data Distribution Service (DDS), a middleware designed for real-time systems that require interoperability, dependability, and scalability. Among various improvements, ROS 2 provides policy-based encryption, authentication, and access control features. However, the authors highlight that ROS 2 security can still be improved, proposing SROS 2, a series of developer tools that add security in complex ROS 2 graphs~\cite{mayoral2022sros2}. The authors discuss the trade-offs and limitations of SROS 2, and they propose a six-steps methodology based on DevSecOps for securing systematically communication graphs. Last, Goerke et al.~\cite{goerke2021controls} noted that, while many security mechanisms and frameworks have been proposed for ROS and ROS 2, none of the current proposals fulfil all the security requirements identified by the authors for securing different use cases. 

Beyond ROS and ROS 2 vulnerabilities, many different attacks on specific sensors and actuators could affect robotic systems operations. As one of many examples, the literature has shown that malicious third parties can strike various spoofing attacks on LiDAR sensors, commonly used in modern robots for navigation purposes. Attackers can spoof LiDARs and prevent them from detecting targets or perceiving non-existing objects, both using digital attack vectors~\cite{cao2021} and physical ones~\cite{jin2023}. Attackers can also fool a LiDAR into misrecognising a moving object in the wrong location by injecting adversarial examples into their learning models~\cite{xu2023sok}.
In conclusion, the current literature shows that even traditional cybersecurity aspects have not been properly addressed in ROS and ROS 2, the most popular frameworks. In light of this, we state our case for the need to take a step further and not only strengthen robots cybersecurity assurances, but also include embodiment in the broader considerations. To the best of our knowledge, this has not been addressed in the literature before.

\subsection{Literature Synthesis}\label{subsec:litsynth}
As aforementioned, while robotics frameworks offer significant modularity and adaptability, they also exhibit critical cybersecurity vulnerabilities. In parallel, studies on cyber-physical systems (CPS) security highlight that attacks are not confined to data breaches or communication disruptions, but can have direct physical consequences. For instance, the compromise of sensors and actuators (e.g., LiDARs) may both hinder task completion and lead to embodiment disruptions that can endanger both the robot and its surroundings. These findings underscore the challenges of ensuring cybersecurity in dynamic and heterogeneous robotic systems. By synthesising these insights, we identified two core cybersecurity challenges in modular robotic systems:
\begin{itemize}
    \item \textbf{Assessing and tolerating disruptions} Robots must evaluate whether a disruption is critical enough to halt operations, or if the system can tolerate it without compromising task performance and physical integrity;
    \item \textbf{Detecting and mitigating threats} Robots need to detect ongoing attacks and choose a mitigation strategy, depending on the potential outcomes.
\end{itemize}

\section{Methodology}
\begin{figure*}[tb]
  \centering \includegraphics[width=\linewidth]{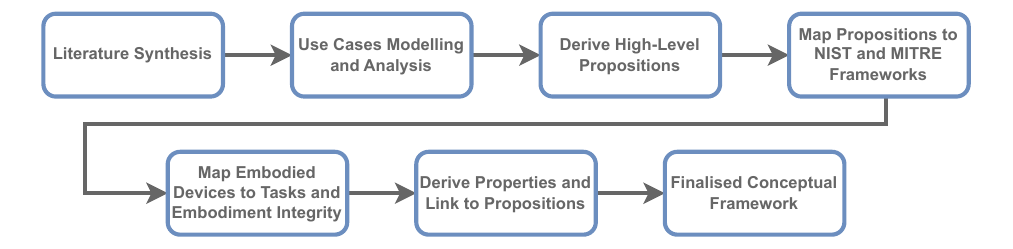}
  \caption{The methodology used in this paper to derive the building blocks of our conceptual framework: propositions and properties.}
  \label{fig:method}
\end{figure*}

The goal of this paper is to provide a conceptual framework to reason on the fundamental characteristics an embodied robot should be equipped with, for dealing with cybersecurity attacks that threaten its embodiment. To address this, we adopt a structured methodology, described in \Cref{fig:method}. Our approach begins with the literature synthesis provided in \Cref{subsec:litsynth}, which summarises the most pressing issues. We follow this with a threat modelling (\Cref{sec:scenarios}), providing two hypothetical use cases that show the challenges an embodied robot might encounter. From these observations, we derive two high-level propositions, and we link them to well-established cybersecurity frameworks (i.e., the NIST Cybersecurity Framework~\cite{nist2024} and the MITRE ATT\&CK for Industrial Control Systems~\cite{mitre2024}). Then, we further analyse our use cases and perform a functional decomposition to identify critical components, in relation to task completion and embodiment integrity preservation. Finally, we use this analysis to derive three final properties, which are the technical mechanisms that allow for implementing our two high-level propositions. This methodology provides a solid foundation for our conceptual framework, which enhances robots cybersecurity and embodiment integrity preservation.

\section{Use Case Analysis - Cyberattacks Consequences on Task Completion and Embodiment}
\label{sec:scenarios}
As previously mentioned in \Cref{sec:intro}, a modern embodied robot can be defined as a set of sensors and actuators (hereby also called devices) that communicate with each other to perform one or multiple tasks, combined in a single physical robot. 
Although these devices are modular and loosely coupled to each other, cyberattacks aimed at devices of an embodied robot might affect the embodiment in different ways and degrees. The effects can range from simply impeding task completion to damaging the entire structural integrity. In this section, we exemplify this idea by proposing two use cases, and we analyse the potential effects of cybersecurity attacks targeted against different devices.

\begin{scen}\label{scen_a}
Let us assume a simple robot, equipped with four independent wheels to move in the environment, one camera to perceive the environment, and a robotic arm to collect objects. The objective of the robot is to navigate in the environment, detect a cube, pick it up, and drop it at a preselected destination.
\end{scen}
\begin{att}\label{att_a1}
In this attack, a malicious party infiltrates the robotic arm and deactivates it.
\end{att}
\begin{an}
This specific cyberattack does not have any repercussions on the integrity of the overall embodiment. The robot can move around the environment as programmed, without any hiccups or increased risks of collisions. On the other hand, the attack disables the robotic arm, and the robot cannot complete the task of picking up the object and delivering it to the destination point. Therefore, the cyberattack has a direct effect on some of the robot tasks. This seems trivial to us, as we have ingrained the arm archetype, its common functions and purposes, and we can derive the implications of a malfunction on the object pick-up task. Robotic systems need explicit data structures and routines, to reason upon these aspects and draw similar conclusions. 
\end{an}

\begin{att}\label{att_a2}
In this attack, a malicious party tampers with the motor controllers of the wheels.
\end{att}
\begin{an}
From the point of view of embodiment integrity, a malfunctioning wheel could lead to unexpected collisions. However, the robot is equipped with a camera which ensures that the robot perceives potential crashes, reducing the chances of collision. However, this is not a guarantee. Depending on the type of disruption (e.g., stopping a wheel or making it work at the fastest speed) and the number of wheel motor controllers corrupted by the attack, the disruption might be easy or hard to mitigate.
\end{an}

\begin{scen}\label{scen_b}
Let us assume a complex humanoid robot that walks on two legs, equipped with two LiDARs for navigating the environment, and a gyroscope to maintain balance. Its goal is to patrol a corridor while maintaining a safe distance from obstacles.
\end{scen}
\begin{att}\label{att_b1}
In this attack, a malicious party tampers with the gyroscope.
\end{att}
\begin{an}
This use case exemplifies a situation where, by compromising a single device, an attacker can threaten both the completion of the goals and the structural integrity of a robot. The gyroscope allows the robot to coordinate the legs to achieve movement and undertake the patrolling task, but also allows to maintain balance (whether while moving or standing still), avoiding collapsing on the ground. In this situation, to avoid fatal collisions, the robotic framework needs to know the role played by the gyroscope in the scheme of the high-level goals, as well as its criticality.
\end{an}
\begin{att}\label{att_b2}
In this attack, a malicious party tampers with one LiDAR, or both.
\end{att}
\begin{an}
In the case of this walking robot, tampering with one of the LiDARs could be a critical threat to the embodiment, as it is one of the two onboard devices that allow it to sense the environment. The robot relies on the LiDARs to navigate the environment, which means that the system relies on the LiDARs to avoid crashes and preserve the integrity of the embodiment. Manipulating one of the two LiDARs might produce wrong information (e.g., altered distance from the objects in the surroundings), causing the robot to crash and be severely damaged. Depending on the degree of corruption of the LiDAR, and if both or only one of them is affected, the attack could have great a influence on basic robot movements and navigation tasks.
\end{an}

\section{Propositions}
\label{sec:propositions}
In the previous section, we described the use cases of two different embodied robots, composed of various devices. We analysed how cyberattacks targeting specific devices can lead to a range of consequences, which vary depending on each component roles and correlated tasks. Building upon the insights derived from the literature synthesis in \Cref{subsec:litsynth}, it becomes clear that a robust cybersecurity framework for modular robots must address both the ability to assess and tolerate disruptions, as well as the capacity to detect and mitigate threats effectively. In this section, we propose two fundamental propositions that outline the resilience and mitigation strategies required for embodied robotic systems to react to cyberthreats. Let us define the first proposition.

\begin{prop}\label{prop_1}
    A robotic framework should be aware of how much disruption and degradation it can tolerate.
\end{prop}
Cyberattacks can use various strategies to disrupt different targets. Some attacks might fully disrupt the target's operations, others might create partial, but dangerous, degradation. When a full disruption occurs on a critical device with no replicas or failback strategies, it is easy to infer that the device has a high criticality score. In \Cref{att_a1}, the robot is equipped with only one arm, meaning that a total disruption due to a cyberattack would render the robot incapable of executing tasks that involve that device, such as pick-and-place of objects. Other cyberattacks affect goals oriented to embodiment preservation. In \Cref{att_b1}, if the gyroscope is corrupted and deactivated by an attacker, the robot loses balance and falls on the ground, threatening its embodiment integrity. 
While these use cases are clear, other cases require a more in-depth analysis. In \Cref{att_b2}, \Cref{scen_b} robot is equipped with two LiDARs. A cyberattack aimed at one of the LiDARs might be negligible, if the framework uses two LiDARs for redundancy purposes, but it has been programmed to work by-design with one LiDAR. On the contrary, if the system has been designed to rely at all times on both LiDARs, one disrupted sensor cannot be tolerated. The ability to evaluate how much a disruption affects a robot in its entirety, which we call \textit{tolerable disruption}, is a building block for our conceptual framework.
 
In other cases, cyberattacks may not cause an immediate failure, but gradually degrade performance. Using again \Cref{scen_b} as an example, the robot relies on timely status updates from the gyroscope, in order to correct its movements in real time and maintain balance. In \Cref{att_b1}, instead of deactivating the gyroscope, an attacker could strike a lightweight Denial of Service (DoS), slowing down the frequency of balance information, just enough to cause the robot to overbalance. The goal is to cause the robot to lose balance and crash while, at the same time, making the attack harder to detect. The ability to sustain partial degradation, which we term \textit{tolerable degradation}, is as important as tolerating complete disruption. A robotic framework must be self-aware of both characteristics, to ensure operational and embodiment integrity. In \Cref{sec:framework}, we show how \Cref{prope_1} and \Cref{prope_2} implement this proposition.

\begin{prop}\label{prop_2}
A robotic framework should be aware of how much disruption and degradation it can mitigate and which countermeasures it should deploy to do so.
\end{prop}

Oftentimes, multiple devices cooperate in task completion and embodiment preservation tasks. Let us refer to \Cref{scen_a} - \Cref{att_a2}, where an attacker deactivates one of the four wheel controllers, in an attempt to steer the robot. Following \Cref{prop_1}, two things can happen. Either the framework is aware that one disrupted wheel does not affect the task execution, and continues its operations. Or it is aware that the disruption cannot be ignored as it affects the tasks or the embodiment preservation goals. In the second case, the framework should be able to react and deploy the available mitigations.

If the framework has reconfiguration capabilities, the robot might be able to carry on with its tasks and preserve the embodiment integrity, even if degradation occurs. Taking as an example \Cref{scen_a} - \Cref{att_a2}, an attacker could make one wheel spin faster than the others, to steer the robot and cause a crash. A proactive framework should be able to compensate for the attack by reconfiguring the other wheel motor controllers, while the system tries to isolate and sanitise the corrupted wheel. One reconfiguration option would be to make the opposite wheel spin equally fast, to match the speed of the tampered wheel and preserve the planned trajectory.
If such mitigation is not applicable, a more decisive intervention could be deployed, such as stopping an ongoing task, such as a pick-and-place. While this option is less desirable than a reconfiguration, as it entails that the pick-and-place cannot be performed, it would cause stopping the three other wheels. As a result, the robot would spin in place, avoiding crashing into objects or people, and preserving its embodiment integrity. Additional steps taken might be the isolation of the tampered devices from other devices, to prevent network lateral movements and further system corruption.

In conclusion, \Cref{prop_2} highlights the need to equip robotic frameworks with awareness of their degrees of \textit{tolerable degradation} and \textit{mitigable degradation}, as defined by \Cref{prop_1}. Without self-awareness of what can be tolerated, it is impossible to determine what can be mitigated. To make matters more complicated, a framework should not only consider disruptions and degradations in isolation. It must also consider their combined effects, in case multiple devices are simultaneously affected by cyberattacks. In \Cref{sec:framework}, we show how \Cref{prope_3} implements this proposition.

\subsection{Alignment with Existing Cybersecurity Frameworks}\label{subsec:alignment}
Before we discuss in our next section how we can fill the technical gap between our propositions and robotic applications, it is necessary to ground the intended outcomes of our propositions in well-established cybersecurity frameworks. In particular, with the help of \Cref{tab:propositions_nist_mitre}, let us discuss how our two propositions align with two established cybersecurity frameworks, the NIST Cybersecurity Framework~\cite{nist2024}, and the MITRE ATT\&CK for Industrial Control Systems (ICS)~\cite{mitre2024}.
\begin{table}[t]
    \centering
    \caption{Correlation between propositions and established cybersecurity frameworks.}
    \label{tab:propositions_nist_mitre}
    \renewcommand\arraystretch{1.3}
    \footnotesize
    \begin{tabular}{P{8em}P{14em}P{14em}} \\
    \textbf{Proposition} & \textbf{NIST Principles} & \textbf{MITRE Tactic}\\ \toprule
    \Cref{prop_1} - Assess and Evaluate  & Identify \& Protect - Identify critical assets and evaluate tolerable degradation and tolerable disruption &  Impact - Recognise consequences of cyberattacks on tasks and embodiment integrity \\
    \Cref{prop_2} - Detect and Mitigate  & Detect \& Respond - Detect threats and attacks, and deploy appropriate mitigation strategies  &  Lateral Movement - Prevent lateral movements within the robotic system and deploy sanitisation strategies\\
    \bottomrule
    \end{tabular}
\end{table}

Starting with NIST, the framework is organised into 6 main categories for managing risks and threats. Of them, 4 are of our interest:
\begin{itemize}
    \item \textbf{Identify} grants that cybersecurity risks are understood;
    \item \textbf{Protect} lists the safeguards in place to minimise risks;
    \item \textbf{Detect} includes the processes to identify and analyse ongoing attacks;
    \item \textbf{Recover} describes solutions to issues identified during the Detect phase.
\end{itemize}

\Cref{prop_1}, aimed at understanding which level of disruption and degradation is tolerable, correlates with the goals of NIST phases Identify and Protect. Similarly, \Cref{prop_2} focuses on processes in place to mitigate cyberattacks and minimise undesired effects on the robotic system, fulfilling the requirements described by the NIST phases Detect and Recovery.

Although the MITRE ATT\&CK focuses on ICS systems, not robotics, there are many similarities between the two fields, taking into consideration the common deployment of robots within industrial settings and mission-critical environments. In particular, the MITRE ATT\&CK lists two general tactics: the Impact and the Lateral Movement. Impact lists possible goals of cyberattacks against critical components, and its categories include damage to property, loss of safety, and denial/loss/manipulation of control. This tactic correlates to our \Cref{prop_1}, providing us with a practical benchmark to evaluate the effects of cyberattacks on our robots. The second tactic, Lateral Movement, describes different techniques that can be used by an attacker to move its presence from one embodied device to others. In particular, the lateral tool transfer sub-tactic shows how it is critical to prevent adversaries from transferring their presence from one embodied device to the other.

\section{Conceptual Framework}
\label{sec:framework}
\begin{figure*}[tb]
  \centering \includegraphics[width=\linewidth]{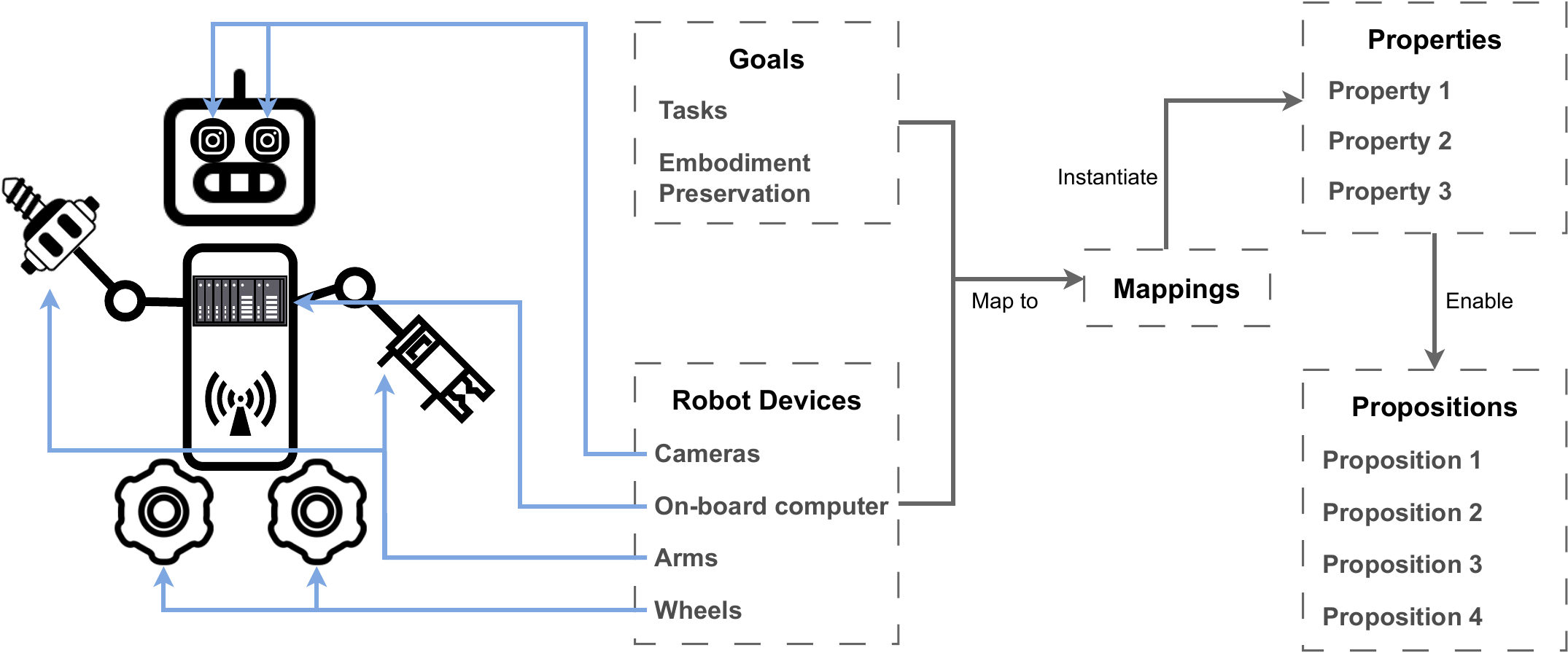}
  \caption{A conceptual robotic framework that shows how the different modules and entities can be composed to allow a robot to achieve the two propositions previously formulated. In particular, the mappings allow to map devices to higher-level goals, hence instantiate the three properties enunciated in this paper.}
  \label{fig:framework}
\end{figure*}

In the previous section, we have provided two propositions, aligned with NIST and MITRE principles, that define high-level security goals. Such propositions enable robotic frameworks to detect and mitigate the effects of cyberattacks, but first we need to define the technical mechanisms that allow us to implement such propositions. In this section, following the conceptual framework shown in \Cref{fig:framework}, we formulate such mechanisms in the form of three properties. We derive the properties from the analysis provided in \Cref{subsec:alignment}, showing that they address the requirements described by the propositions and by the cybersecurity frameworks created by NIST and MITRE.

\subsection{Property 1}
Let us start with formulating the first property. \Cref{prop_1} puts emphasis on the ability to assess robotic systems tolerance to disruptions, and it aligns with the NIST Identification and Protection principles. From this observation, we derive the following property:
\begin{prope}\label{prope_1}
    \textbf{Device Task-Criticality} A robotic framework should be able to assess device role and criticality for achieving task completion.
\end{prope}
This property ensures the framework identifies which devices are critical, for the execution of the expected tasks. It is the simplest of the three properties, but it is fundamental for enabling the aforementioned propositions, hence the other properties. For example, \Cref{prop_1} states that a robot should be able to stop tasks that require the collaboration of a device disrupted by a cyberattack, assuming that the robot is aware of which devices are involved in each task. Therefore, a list of devices, a list of tasks, and a mapping between the two lists is necessary. Referring to \Cref{scen_a}, the framework should be equipped with a mapping similar to the one shown in \Cref{tab:1}. It is worth noting that mapping if a device is involved in a task, or not at all, is straightforward; much less straightforward is determining to which extent the device is critical for completing successfully a task. Therefore, defining a mapping like the one shown in \Cref{tab:1} may not be trivial. In particular, in \Cref{tab:1} we evaluated the wheels as important devices for moving, but not critical, while we marked the camera critical. This might sound counter-intuitive, but this signals that for this specific embodied robot, one single disrupted wheel motor controller is not going to entirely prevent movement if the other three wheels are still operating. On the other hand, the robot is equipped with only one camera for sensing the environment: if that is disrupted, the robot loses any perception capability.

\Cref{prope_1} is necessary to allow a robotic framework to achieve \Cref{prop_1}, for what it concerns task completion. In the first case, the framework can evaluate if tasks can be carried on if a device is disrupted, only when it is aware if (and to what extent) the device is necessary for the tasks. Similarly, without a mapping between devices and tasks, it would be impossible for the framework to evaluate which tasks are affected by a corrupted device. Consequently, it cannot evaluate which other devices it should stop, either.
This mapping could be provided in the form of an explicit list, but this approach lacks the flexibility of adapting to embodiment dynamic changes (e.g., a sensor swapped with another one). In alternative, this mapping could be inferable by executing machine learning (ML) routines that take as inputs the devices metadata, the services offered by the devices, and the list of tasks. As output, these ML-driven routines would produce the mapping. An example of machine learning routines that allow a similar result is presented in the $S\times C$ IoT framework~\cite{giaretta2021s}, which exhibits the capability of inferring IoT devices type and role by using their metadata.

\begin{table}[t]
    \centering
    \caption{Example of a mapping between devices and their involvement in tasks completion, to achieve Proposition 1. Symbols: (\X) device critical for task completion. (\til) device important for task completion. (-) device not involved in the task.}
    \label{tab:1}
    \renewcommand\arraystretch{1.3}
    \footnotesize
    \begin{tabular}{lP{5em}P{5em}P{5em}P{5em}}
    & \multicolumn{4}{c}{\textbf{Task}} \\
    \cmidrule(lr){2-5}
    \textbf{Device} & Move & Find Object & Pick Object & Drop Object\\ \toprule
    Wheel 1     & \til  &  -    &  -    &  -    \\
    Wheel 2     & \til  &  -    &  -    &  -    \\
    Wheel 3     & \til  &  -    &  -    &  -    \\
    Wheel 4     & \til  &  -    &  -    &  -    \\
    Camera      & \X     & \X    & -     &  -    \\
    Robotic Arm &  -     &  -    & \X    & \X    \\
    \bottomrule
    \end{tabular}
    
\end{table}

\subsection{Property 2}
Building on the need to recognise the consequences of cyberattacks on embodiment integrity, highlighted by \Cref{prop_1} and supported by the MITRE Impact tactic, we derive the following:
\begin{prope}\label{prope_2}
    \textbf{Device Embodiment-Criticality} A robotic framework should be able to assess the role and criticality of the device in preserving the structural integrity of the embodiment.
\end{prope}
This property focuses on evaluating the importance of each embodied device for the preservation of the embodiment structural integrity. \Cref{prope_1}, as previously defined, allows us to know which tasks could be affected in case of ongoing cyberattacks targeted to specific devices. Similarly, \Cref{prope_2} concerns with the effect of cyberattacks (targeted at specific devices) on the entire structural integrity. In \Cref{prope_1}, we map devices to tasks execution, such as moving to destination or pick and place tasks (e.g., as shown in \Cref{scen_a}. Achieving \Cref{prope_2} requires a similar mapping effort, but this time in the form of evaluating the effects of cyberattacks on the entire embodiment. In its simplest form, this could be obtained again with a table similar to \Cref{tab:1} shown for \Cref{prope_1}. In this case, it is necessary to map devices to their importance for achieving goals that concern with the preservation of the embodied robot. This could be expressed with a list of high-level preservation principles that guide the robot. For example, referring to \Cref{scen_a}, the robot could be provided with two basic principles, such as avoiding falling down stairs and avoiding bumping into obstacles, as shown in \Cref{tab:2}.

\begin{table}[t]
    \centering
    \caption{Example of a mapping between devices and their involvement in embodiment preservation to achieve Proposition 1. Symbols: (\X) device critical for fulfilling the goal. (\til) device important for fulfilling the goal. (-) device not involved in goal fulfilment.}
    \label{tab:2}
    \renewcommand\arraystretch{1.3}
    \footnotesize
    \begin{tabular}{lP{8em}P{8em}}
    & \multicolumn{2}{c}{\textbf{Preservation Principle}} \\
\cmidrule(lr){2-3}
    \textbf{Device} & Avoid Obstacles & Avoid Stairs \\ \toprule
    Wheel 1     & \til  & \til  \\
    Wheel 2     & \til  & \til  \\ 
    Wheel 3     & \til  & \til  \\ 
    Wheel 4     & \til  & \til  \\ 
    Camera      & \X    & \X    \\
    Robotic Arm & -     & -     \\
    \bottomrule
    \end{tabular}
    
\end{table}

\Cref{prope_2} plays an important role in the preservation of embodiment integrity, in all propositions \Cref{prop_1,prop_2}. Indeed, without a clear mapping between devices and their role in guaranteeing a robust embodiment, it is not possible for a framework to evaluate which and how many disrupted devices can be tolerated, how much service degradation can be tolerated or mitigated, nor when certain tasks should be preemptively stopped to avoid fatal crashes.
As mentioned previously, while this could be achieved with an explicit list of preservation principles, this approach does not allow the framework to react to embodiment dynamic changes. Again, the same mapping could be achieved by ML routines, although the task is more complicated than for \Cref{prope_1}. In this case, the routines would have to list the devices installed on the embodiment, derive which kind of embodiment has been achieved by combining the devices, and infer which threats could hit the embodiment.

Let us exemplify this by comparing \Cref{scen_a} and \Cref{scen_b}. In the first case, the ML-driven routines would realise that the robot has four wheels, a camera, and no gyroscope. From this information, the framework would infer that the robot is four-wheeled, that is expected to avoid obstacles, but is not expected to encounter situations that would make it tip over (hence, no need for the gyroscope). Therefore, the framework assumes that the robot needs to avoid obstacles and falling down stairs (or similar situations), and it links the preservation principles to the devices that might have a role. In the second case, the robot has two legs, two LiDARs, and a gyroscope. Based on these devices, the routines infer that the robot must also maintain balance, in addition to avoiding obstacles and falling down staircases. While this is easy for human beings, due to their ability of building mental archetypes from past experience, it is far from a trivial task to implement such routines in a robot. These routines, that could be ML-based as the ones hypothesised in \Cref{prope_1}, must infer the robot archetype (e.g., walking robot or four-wheeled robot), based on the installed devices and their possible relation to each other within the specific embodiment. Once the archetype is known, higher preservation principles could be inferred.

\subsection{Property 3}
Some devices, although not directly critical for the embodiment preservation, can serve as entry points for tampering with other embodied devices. To prevent these network lateral attacks, highlighted by \Cref{prop_2} and supported by the MITRE Lateral Movement tactic, we derive the following property:
\begin{prope}\label{prope_3}
    \textbf{Inter-Device Awareness} A robotic framework should be capable of assessing the criticality of the device for the embodiment structural integrity, based on the possibility to use it to attack other embodied devices.
\end{prope}
An additional level of complexity stems from the consideration that devices must interact and communicate with each other, to compose an embodiment. Embodied devices can be loosely or tightly coupled, depending on goals and functions, but every devices is somewhat coupled to another device, at least. In other words, embodied devices communicate with other embodied devices, and establish information flows. This property evaluates a device potential to compromise other components, thereby influencing the overall embodiment integrity.

While coupling is necessary for achieving a working embodiment, it entails higher cybersecurity risks. An attacker could penetrate a non-critical embodied device and use it as an entry point for penetrating other devices, critical for the robot integrity. Therefore, although a vulnerable device might not be a direct threat for the embodiment integrity, it should be considered equally critical if it can be used as an entry point for compromising a critical device. The more connected and coupled the devices, the higher the risks. This attack strategy, known as network lateral movement, showcases the importance of evaluating how embodied devices communicate with each other and if events could be chained to perform network lateral movements. 

In addition, actions in the physical domain could be chained to achieve unexpected results, as shown in the literature by Agadakos et al.~\cite{agadakos2017}. For example, in \Cref{scen_a}, depending on where the robotic arm has been installed, an attacker could use it for obstructing on purpose the camera field of vision. Therefore, a framework should be able to model physical interactions as hidden communication channels and their effects on tasks, as discussed by Giaretta et al.~\cite{giaretta2019}, as well as on embodiment preservation goals.

\begin{table*}[t]
    \centering
    \caption{Example of a mapping that implements Proposition 2. Each device is mapped to three entities: the devices with which it exchanges information, the probability of being subject to a cybersecurity attack, and its criticality in ensuring embodiment preservation. The last column presents for each device the overall criticality score, in the scope of the embodiment preservation. For this example, we assign Low for scores lower than 0.5, Medium for scores lower than 1.5, and High for scores higher than 2.}
    \label{tab:3}
    \renewcommand\arraystretch{1.3}
    \footnotesize
    \begin{tabular}{lP{6em}P{8em}P{8em}P{8em}}
    & & \multicolumn{2}{c}{\textbf{Scores}} & \\
\cmidrule(lr){3-4}
    \textbf{Device} & \textbf{Connected to Devices} & \textbf{Vulnerability Probability (\%)} & \textbf{Criticality for Embodiment (1-3)}  & \textbf{Overall Criticality for Embodiment} \\ \toprule
    Wheel 1     & Wheel 2, Wheel 3, Wheel 4 & \zerotwo  & \medium     & \overallmedium \\
    Wheel 2     & Wheel 1, Wheel 3, Wheel 4 & \zerotwo   & \medium      & \overallmedium \\
    Wheel 3     & Wheel 1, Wheel 2, Wheel 4 & \zerotwo   & \medium      & \overallmedium \\
    Wheel 4     & Wheel 1, Wheel 2, Wheel 3 & \zerotwo   & \medium     & \overallmedium \\
    Camera      & Robotic Arm               & \zerofive     & \high       & \overallhigh \\
    Robotic Arm & Camera                    & \zeronine   & \low        & \overallhigh \\
    \bottomrule
    \end{tabular}
    
\end{table*}

A strategy for achieving this third property, is to equip the framework with a systematic process to evaluate how much a cyberattack targeted on the device would affect the entire embodiment. This can be done by combining the analysis of the stand-alone cybersecurity for every device and their degree of connectivity with other embodied devices, using predefined formulas or ML approaches. The result would look similar to what we show in our toy example in \Cref{tab:3}, where we see that the framework scores each device on its cyber-physical vulnerabilities, its criticality for preserving the embodiment, and which devices it communicates directly with. As a result, the framework outputs an overall score for each device, that indicates how critical it is to harden its cybersecurity properties. These insights could lead to re-evaluate the criticality score for some devices. For example, as shown in \Cref{tab:3}, the robotic arm has low criticality, as an isolated device. However, the device is very vulnerable (high vulnerability score) and it communicates with the camera. Based on the information that the camera is critical for preserving the embodiment, and that it has a medium chance of being vulnerable to cyberattacks, the framework highlights that the robotic arm could be used as an initial intrusion point for performing a network lateral movement. Therefore, despite its low criticality score for the embodiment (as an isolated device), the arm is considered a device to protect with priority. 

The scores assigned to \textit{vulnerability probability} and \textit{criticality for embodiment} in \Cref{tab:3} were picked arbitrarily, as in reality they would depend on the specific device at hand. For the \textit{overall criticality}, we have computed the scores following \Cref{eq:overall} described in \Cref{subsec:devcrit}, and we have arbitrarily mapped the numerical scores to a \textit{Low} - \textit{High} range. \Cref{eq:overall} , detailed in \Cref{subsec:devcrit}, illustrates a potential approach for evaluation. While not central to the core contribution of this paper, it serves as an example of how different evaluation methods can be structured depending on specific use case requirements.

\Cref{prope_3} is necessary to achieve \Cref{prop_1,prop_2}, as it evaluates a device criticality score based on the possibility of using it to affect other devices, and their own criticality for the embodiment (determined by \Cref{prope_2}). All three \Cref{prope_1,prope_2,prope_3} are necessary for implementing the necessary routines that achieve the NIST principles Detect and Respond. As shown throughout this section, our propositions and properties are the foundational blocks that allow a robotic system to achieve two goals. First, detecting potential threats and ongoing cyberattacks, on both embodiment tasks and integrity. Second, deploying appropriate mitigation strategies for responding to ongoing cyberattacks.

\section{Future Work}\label{sec:futwok}
In this paper, we have laid the foundations of a conceptual framework, but various aspects need to be explored further. First, we must implement our framework to evaluate its strengths and weaknesses. Acquiring different devices and building robots is a costly task. Therefore, our plan is to implement our framework using a robotic simulator, such as Gazebo~\cite{gazebo} or CoppeliaSim~\cite{coppeliasim}. These simulators will allow us to test our framework on diverse robots, composed with different embodied devices, in a variety of physical environments, and under different cyberattacks. In simulation, it is possible to simulate different cyberattacks, from DoS attacks that increase packets loss and jitter, to total disruption of one or more embodied devices at once. Eventually, we will follow our simulations with an implementation on real robotic systems, for evaluating the transferability of our framework from simulation to the real world.

Moreover, the process of mapping devices to their roles in task completion and embodiment preservation is central to the framework. In this first paper, we offer illustrative examples to showcase the need for such a framework, and we note that the criteria for determining what makes a device critical is context-dependent. For instance, in some cases we could evaluate this based on comprehensive historical data on vulnerabilities and exploits, while in other situations such data may be limited or not available. At the conceptual stage of this paper, it is not feasible to make universal assumptions on this regard. In our future work, we will develop a rigorous, adaptable methodology for producing this mapping, potentially leveraging statistical modeling, machine learning techniques, or expert elicitation to accommodate varying data availability and application scenarios.

\section{Conclusion}
\label{sec:conclusion}
Various traditional cybersecurity issues have been identified by the literature in popular robotics middleware, ROS and ROS 2. In this paper, we claim that not only it is important for the robotics field to improve on these aspects. 
It is also critical to achieve more substantial advances, concerning the correlation between robotic embodiments, their composition and integrity, and their cybersecurity properties. To substantiate our claims, we have described two use cases of embodied robots and two cybersecurity attacks for each use case. Then, we showed how cyberattacks targeted at different devices can have different consequences, depending on the nature of the targeted devices, but also on their role within the embodiment. We provided an in-depth analysis of such cyberattacks, and formalised in two propositions the characteristics that robotic frameworks should exhibit to mitigate the connected threats. The propositions have been grounded on two foundational cybersecurity frameworks, the NIST Cybersecurity Framework~\cite{nist2024} and the MITRE ATT\&CK for ICS~\cite{mitre2024}. Finally, as a technical mechanism to implement the propositions, we have provided three properties, with related components and routines.

\section*{Acknowledgement}
This work was partially supported by the Wallenberg AI, Autonomous Systems and Software Program (WASP) funded by the Knut and Alice Wallenberg Foundation, and by the Wallenberg AI, Autonomous Systems and Software Program - Humanities and Society (WASP-HS) funded by the Marianne and Marcus Wallenberg Foundation and the Marcus and Amalia Wallenberg Foundation.

\bibliographystyle{splncs04}
\bibliography{aaai24}

\appendix
\section{Appendix - Device Criticality Equation}\label{subsec:devcrit}
For the scope of this paper, we do not need a detailed model for calculating overall criticality. We need a general one that demonstrates the feasibility of our approach, but that is also flexible enough for tuning, as different scenarios might need to weigh in information differently.

Let us define three fundamental components. The first component is the degree of connectivity \textit{n}, which is the number of devices directly connected to the device currently analysed. Then, we have the likelihood that an attacker leverages a vulnerability to penetrate the device \textit{V}; this probability is based on past incidents, CVEs, and CWEs, and is expressed within a $[0,1]$ range. Last, we take into consideration how critical is the device for the integrity of the embodiment, indicated with the term \textit{E}, and expressed in a 3-point scale of \textit{Low}, \textit{Medium}, and \textit{High}. As an example, a robot relying only on one camera for navigating the environment would certainly crash without a working camera, hence the camera would have a \textit{High} score for \textit{E}. These components are all combined to provide an overall criticality score $O_D$, following the equation:
\begin{equation}\label{eq:overall}
    O_D =  \alpha(V_D\cdot E_D)+\beta\frac{1}{n}\sum_{i=0}^{n}(V_i\cdot E_i).
\end{equation}
In the second portion of this equation, we evaluate the average impact of all the neighbouring devices, instead of the total, to avoid situations in which a device criticality is overestimated just because of its high connectivity. In addition, we have introduced two parameters $\alpha$ and $\beta$, which allow to tune the weight assigned to the device itself, and the weight assigned to the connected neighbouring devices, respectively. For the sake of the example covered in \Cref{prope_3}, we have used a simplified formula with $\alpha=1$ and $\beta=1$.

As discussed in \Cref{sec:rel_work}, compromised devices have the potential to be chained to perform several lateral network movements. With \Cref{eq:overall}, we provide a general equation for evaluating the overall impact of insecure devices on embodied robotic systems, taking into account the degree of connectivity of each device, hence the risk of lateral network movements. However, we have taken into consideration only the first degree of neighbours. Our framework would benefit from the capability of evaluating chained paths that could result in multiple-rounds lateral network movements. Besides, our equation uses different parameters that can be tuned to fit the model to the robotic system at hand, but it does not explore how tuning strategies affect the outcome of the model. In our future work, we will delve into these aspects, besides the ones covered in \Cref{sec:futwok}.

\end{document}